\documentclass[prc,amsmath,superscriptaddress]{revtex4}

\usepackage{latexsym}
\usepackage{amssymb}
\usepackage{dsfont}
\usepackage{amsbsy}
\usepackage{amsmath}
\usepackage[varg]{txfonts}
\usepackage{mathrsfs}
\usepackage{upgreek}
\usepackage{verbatim}
\usepackage{hyperref}
\newsavebox{\fmbox}

\DeclareMathAlphabet{\mathpzc}{OT1}{pzc}{m}{it}

\usepackage{graphicx,color}
\usepackage{amssymb}
\usepackage{enumerate}
\usepackage{verbatim}
\usepackage{natbib}
\usepackage{color}
\usepackage{array}
\begin{document} 
\title{Inclusive deuteron--induced reactions and final neutron states}

\author{G. Potel}
\affiliation{National Superconducting Cyclotron Laboratory, Michigan State University, East Lansing, Michigan 48824, USA}
\affiliation{Lawrence Livermore National Laboratory L-414, Livermore, CA 94551, USA}
\author{F.~M.~Nunes}
\affiliation{National Superconducting Cyclotron Laboratory, Michigan State University, East Lansing, Michigan 48824, USA}
\affiliation{Department of Physics and Astronomy, Michigan State University, East Lansing, MI 48824-1321}
\author{I.J. Thompson}
\affiliation{Lawrence Livermore National Laboratory L-414, Livermore, CA 94551, USA}	
\begin{abstract}
We present in this paper a formalism for deuteron--induced inclusive reactions. We disentangle direct elastic breakup contributions from other processes (which we generically call non--elastic breakup) implying a capture of the neutron both above and below the neutron emission threshold. The reaction is described as a two step process, namely the breakup of the deuteron followed by the propagation of the neutron--target system driven by an optical potential. The final state interaction between the neutron and the target can eventually form an excited compound nucleus. Within this context, the direct neutron transfer to a sharp bound state is a limiting case of the present formalism. 
\end{abstract}
\maketitle 

\section{Introduction}
 The population of discrete neutron states with $(d,p)$ transfer reactions is a well established experimental method. It has proven to be the tool of choice for the study of the single--particle nature of states close to the Fermi energy, providing information about the energy, spin, parity, and spectroscopic factors, of those states. As a result of the coupling with more complex nuclear degrees of freedom, some of them get fragmented and spread over a finite energy region, and, as we move away from the Fermi energy, they acquire larger energy widths. As we go towards the neutron drip line, the Fermi energy gets closer to the neutron--emission threshold, and, eventually,  slides into the continuum. Standard direct transfer reaction theory, such as the Distorted Wave Born Approximation (DWBA) and coupled channels approaches, deal, as a rule, with the population of sharp discrete states. They are thus not well adapted for the description of the transfer to wide states, let alone to states in the continuum region of the spectrum. Early works to provide a more suitable formalism were initiated in the late 70's, but the activity in this field ended quite abruptly in the early 90's, leaving behind an unresolved controversy regarding different approaches (\cite{Kerman:79,Udagawa:80,Austern:81,Li:84,Udagawa:86,Ichimura:85,Ichimura:90,Hussein:85,Hussein:89}). Recently, a few groups have revived the subject, producing different computer codes to implement the reaction formalism (\cite{Potel:15b,Lei:15,Carlson:15}). Though their approaches are slightly different, they are all based on a two--step description of the reaction mechanism. The two steps considered to describe the deuteron--target reaction are the  breakup of the deuteron followed by a propagation of the loose neutron in the target field. This field is modeled with an optical potential, and can account for the absorption of the neutron both in finite--width bound states and in the above neutron--emission threshold continuum states.

  Aside from providing valuable spectroscopic information  about the nature of single-particle states in  nuclei, the absorption of the neutron can be used at profit to study neutron--induced reactions in radio active isotopes with the surrogate reaction method in inverse
 kinematics. A considerable theoretical and experimental effort is being devoted to the study of neutron capture $(n,\gamma)$ and neutron induced fission $(n,f)$ reactions in exotic nuclei making use of the surrogate method (\cite{Escher:10,Scielzo:10,Escher:12,Hughes:12,Ross:12,Carlson:14}). In these experiments, an exotic beam impinging on a deuteron target absorbs the neutron of the deuteron, forming an (as a rule) excited compound nucleus that later decays emitting principally $\gamma$ radiation and neutrons. The theoretical prediction of the cross section for the formation of the compound nucleus in a state of given excitation energy, angular momentum and parity (see \cite{Potel:15b}) is key for the extraction of the  $(n,\gamma)$ cross sections from the analysis of the experiment (\cite{Carlson:14}). In  section (\ref{intro}) we  briefly introduce the formalism,  (we refer to \cite{Potel:15b} for a detailed derivation), and provide specific expressions for the numerical calculation of non--elastic breakup cross sections. In section (\ref{results}) we show examples of final neutron states, and we discuss the relationship between the population of states below neutron--emission threshold and the direct neutron transfer in the DWBA approximation.  
\section{Theoretical formulation}\label{intro}
\label{theory}
\subsection{General formalism in the \emph{prior} representation}
\label{theory-AV}

Let us consider the reaction A(d,p)B* which includes elastic breakup and any other inelastic processes.  
The three-body Hamiltonian for the problem is
\begin{align}\label{eq24}
H=K_n+K_p+h_A(\xi_A)+V_{pn}(r_{pn})+V_{An}(r_{An},\xi_A)+U_{Ap}(r_{Ap}),
\end{align}
where $K_n$ and $K_p$ are the kinetic energy operators acting on the neutron and  proton coordinates respectively.  We have adopted a spectator approximation for the outgoing  proton, we thus model its interaction with the target by means of an optical potential $U_{Ap}$. The coordinates used throughout are defined in Fig. \ref{fig1}.
Starting from neutron--target ($U_{An}$) and deuteron--target ($U_{Ad}$) optical potentials, we can define the optical model Green's function in the breakup channel,
\begin{align}\label{eq31}
G_B^{opt}=\frac{1}{E-E_p{-}\varepsilon_A-K_n-U_{An}(r_{An})+i\epsilon},
\end{align}
and the source term
   \begin{align}\label{eq57}
S_{\rm prior}=\left(\,\chi_f^{(-)} \big |  U_{Ap} - U_{Ad}+U_{An} \big |\phi_d\,\chi_i\right\rangle,
   \end{align}
   where round bracket indicates integration over the proton coordinate only, and 
   the proton distorted wave $\chi_f^{(-)}$ satisfies the equation
   \begin{align}\label{eq140}
   \left(E_f-K_p-U_{Ap}^\dagger\right)\chi_f^{(-)}=0,
   \end{align}
   where $E_f$ is the final channel energy.
   We can then define the neutron final wavefunction in the \textit{prior} representation,
   \begin{align}\label{eq143}
   \psi_{n}^{prior}  =G_B^{opt} S_{\rm prior},
         \end{align}
         and the non-orthogonality function
            \begin{align}\label{eq45}
            \psi_n^{HM}=\left(\chi_f^{(-)}\right|\,\left.\,\vphantom{\chi_f}\phi_d\,\chi_i\right\rangle.
            \end{align}
It can be shown (see \cite{Potel:15b}) that the non--elastic breakup cross section in the \textit{prior} representation can then be written in term of (\ref{eq143}) and (\ref{eq45}) as
   \begin{align}\label{eq58}
  \nonumber & \left.\frac{d^2\sigma}{d\Omega_pdE_p}\right]^{NEB}=-\frac{2}{\hbar v_d}\rho_p(E_p)\left[\Im\left\langle\,\psi_n^{prior}|W_{An}\,|\psi_n^{prior}\right\rangle \right.\\ 
  &  +2\Re\left\langle\,\psi_n^{{HM}}|W_{An}|\psi_n^{prior}\right\rangle +\left.\left\langle\,\psi_n^{{HM}}|W_{An}\,|\psi_n^{{HM}}\right\rangle\right],
   \end{align}
 where
  \begin{align}\label{eq82}
 \rho_p(E_p)=\frac{m_p k_p}{8\pi^3\hbar^2}
  \end{align}
  is the proton level density, and $E_p$ is the kinetic energy of the detected proton.

\subsection{Partial wave expansion}
 \begin{figure}
 \begin{center}
 \includegraphics[width=6cm]{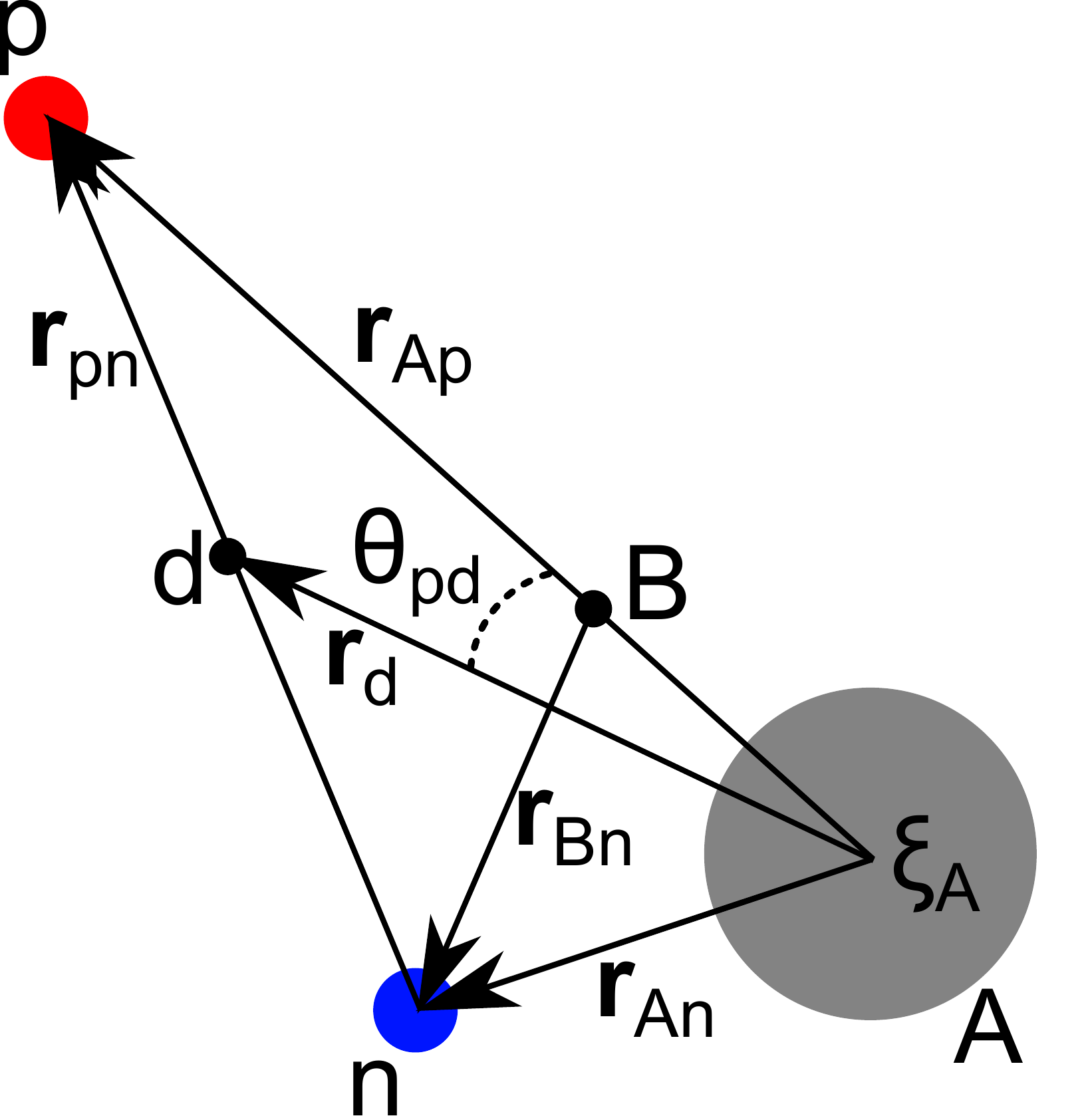}
 \end{center}
\caption{Schematic representation of the system under consideration with the coordinates used in its description.}\label{fig1}
\end{figure}
The implementation of the formalism relies on the numerical evaluation of the source term
   \begin{align}\label{eq90}
\nonumber S_{\text{prior}}(\mathbf{r}_{Bn};\mathbf{k}_p)&=\left\langle\chi_p\right|V\,\left|\,\vphantom{\chi_p}\phi_d\,\chi_d\right\rangle\\
&=\int d \mathbf{r}_{Ap}\, \chi_p^{(-)*}(\mathbf{r}_{Ap};\mathbf{k}_p)V(r_{An},r_{Bn},r_{pn})\phi_d(\mathbf{r}_{pn})\chi_d^{(+)}(\mathbf{r}_{d}).
   \end{align}
It is convenient to express the quantities of interest in terms of a partial wave expansion
\begin{align}\label{eq15}
S_{\text{prior}}(\mathbf{r}_{Bn};\mathbf{k}_p)=\frac{2m_n}{\hbar^2}\sum_{lml_p}\mathcal {F}_{lml_p}(r_{Bn};\mathbf{k}_{p})Y_{m}^l(\theta_{Bn})Y^{l_p}_{-m} (\hat k_{p}).
\end{align}
 Let's first extract the  dependence of the neutron final angular momentum $l$ by defining the $F$ coefficients,
\begin{align}\label{eq16}
{F}_{lm}(r_{Bn};\mathbf{k}_{p})=\int d\Omega_{Bn} S_{\text{prior}}(\mathbf{r}_{Bn};\mathbf{k}_p) Y_{m}^{l*}(\theta_{Bn}),
\end{align}
and  $d\Omega_{Bn}\equiv\sin(\theta_{Bn})\,d\theta_{Bn}\,d\varphi_{Bn}$.
The distorted waves of the proton and the deuteron can be expanded in partial waves in a standard way, 
 \begin{equation}
\chi_p^{(-)*}(\mathbf{r}_{Ap};\mathbf{k}_p)= \frac{ 4\pi }{k_p r_{Ap}}\sum_{l_p} i^{-l_p}
e^{i\sigma_p^{l_p}} f_{l_p}  (r_{Ap}) \sqrt{2l_p+1}\left[ Y^{l_p} (\hat r_{Ap}) Y^{l_p} (\hat k_p)\right]^0_0,
\end{equation}
where $f_l(r_{Ap})$ is the solution, for each partial wave, of the radial part of the Schr\"odinger equation with an optical potential $U_{Ap}(r_{Ap})$.
 \begin{equation}\label{eq11}
\chi_d(+)(\mathbf{r}_{d})= \frac{ 4\pi }{k_d r_{d}}\sum_{l_d} i^{l_d}
e^{i\sigma_d^{l_d}} g_{l_d}  (r_{d}) \sqrt{2l_d+1}\left[ Y^{l_d} (\hat r_{d}) Y^{l_d} (\hat k_d)\right]^0_0.
\end{equation}
In this last expression, $g_l(r_d)$ is the solution, for each partial wave, of the radial part of the Schr\"odinger equation with the optical potential $U_{Ad}(r_{d})$ describing the relative motion between the deuteron and $A$ in the initial channel. 
If we only take into account the $S$--wave component of the deuteron wavefunction, we can write
 \begin{equation}
\phi_d(\mathbf{r}_{pn})=\frac{1}{\sqrt{4\pi}}u_d(r_{pn}).
\end{equation}
Then
 \begin{align}
\nonumber  {F}_{lm}&(r_{Bn};\mathbf{k}_{p})=\frac{ 8\pi^{3/2} }{k_d k_p}\sum_{l_pl_d}i^{l_d-l_p}e^{i(\sigma_p^{l_p}+\sigma_d^{l_d})}\sqrt{(2l_p+1)(2l_d+1)}\\
&\nonumber\times\int d r_{Ap}\,d\Omega_{Ap}\,d\Omega_{Bn}r_{Ap} \frac{f_{l_p}(r_{Ap})\,g_{l_d}(r_d)}{r_d}u_d(r_{pn}) V(r_{An},r_{Bn},r_{pn}) \\
&\times\left[ Y^{l_p} (\hat r_{Ap}) Y^{l_p} (\hat k_p)\right]^0_0 \left[ Y^{l_d} (\hat r_{d}) Y^{l_d} (\hat k_d)\right]^0_0Y^{l*}_m (\theta_{Bn}).
\end{align}
After some Racah algebra, we get
 \begin{align}\label{eq64}
\nonumber  {F}_{lm}&(r_{Bn};\mathbf{k}_{p})=\frac{ 8\pi^{3/2} }{k_d k_p}\sum_{l_d-l_p}i^{l_d-l_p}e^{i(\sigma_p^{l_p}+\sigma_d^{l_d})}\sum_{KM}(-1)^{K-M}\left[ Y^{l_p} (\hat k_{p}) Y^{l_d} (\hat k_{d})\right]^K_{-M}\\
&\nonumber\times\int d r_{Ap}\,d\Omega_{Ap}\,d\Omega_{Bn}r_{Ap} \frac{f_{l_p}(r_{Ap})\,g_{l_d}(r_d)}{r_d}u_d(r_{pn}) V(r_{An},r_{Bn},r_{pn}) \\
&\times\left[ Y^{l_p} (\hat r_{Ap}) Y^{l_d} (\hat r_{d})\right]^K_M (-1)^{l-m}Y^{l}_{-m} (\theta_{Bn}).
\end{align}
 We can then make the replacement
 \begin{align}\label{eq80}
\nonumber \left[ Y^{l_p} (\hat r_{Ap}) Y^{l_d} (\hat r_{d})\right]^K_M &Y^{l}_{-m} (\theta_{Bn})\rightarrow\langle l\;l\;m\;-m|0\;0\rangle\left\{\left[ Y^{l_p} (\hat r_{Ap}) Y^{l_d} (\hat r_{d})\right]^l Y^{l}(\theta_{Bn})\right\}^0_{0}\\
&=\frac{(-1)^{l-m}}{\sqrt{2l+1}}\left\{\left[ Y^{l_p} (\hat r_{Ap}) Y^{l_d} (\hat r_{d})\right]^l Y^{l}(\theta_{Bn})\right\}^0_{0},
\end{align}
as all other possible angular momentum couplings integrate to zero. Note that this is required for angular momentum conservation.
The integrand being rotationally invariant, we can evaluate it for a particular configuration (say, the $z$--axis along $\mathbf{r}_{Bn}, \mathbf{k}_p$ and $\mathbf{r}_{Ap}$ lying in the $xy$ plane) and multiply the result by a factor of $8\pi^2$ (resulting from the integration over $\varphi_{Ap}, \varphi_{Bn}$ and $\theta_{Bn}$). Then one can then write the $l,m,l_p$ coefficient defined in eq. (\ref{eq15}) as the  2--D integral that is numerically evaluated in our code, 
 \begin{align}\label{eq17}
\nonumber \mathcal {F}_{lml_p}&(r_{Bn};\mathbf{k}_{p})=(-1)^m\frac{ 16\pi^{5/2} }{k_d k_p}\sum_{l_d}i^{l_d-l_p}e^{i(\sigma_p^{l_p}+\sigma_d^{l_d})}\langle l_p\;l_d\;-m\;0|l\;-m\rangle \\
&\nonumber\times  \sqrt{\frac{2l_d+1}{2l+1}}\int r_{Ap} d  r_{Ap}\,\sin(\theta) d\theta  \frac{f_{l_p}(r_{Ap})\,g_{l_d}(r_d)}{r_d}\\
&\times u_d(r_{pn}) V(r_{An},r_{Bn},r_{pn}) 
\left[ Y^{l_p} (\theta) Y^{l_d} (\theta_d)\right]^l_0,
\end{align}
where $\theta_d,r_{pn},r_d$ are obtained as functions of $r_{Ap},r_{Bn},\theta, \theta_{Bn}$ according to the definitions found in  Fig. \ref{fig1}, and are to be evaluated for $\theta_{Bn}=0$.

The non orthogonality term defined in eq. (\ref{eq45}) can also be expanded in partial waves in a very similar  way,
 \begin{equation}\label{eq71}
\psi_n^{HM}(\mathbf{r}_{Bn};\mathbf{k}_p)= \sum_{l,m,l_p}\phi^{HM}_{lml_p}(r_{Bn};\mathbf{k}_{p})Y^l_m(\theta_{Bn})Y^{l_p}_{-m} (\hat k_{p})/r_{Bn},
\end{equation}
with
  \begin{align}\label{eq74}
 \nonumber \phi^{HM}_{lml_p}&(r_{Bn};\mathbf{k}_{p})=(-1)^m\frac{ 16\pi^{5/2} }{k_d k_p}\sum_{l_d}i^{l_d-l_p}e^{i(\sigma_p^{l_p}+\sigma_d^{l_d})}\langle l_p\;l_d\;-m\;0|l\;-m\rangle \\
 &\times  \sqrt{\frac{2l_d+1}{2l+1}}\int r_{Ap} d  r_{Ap}\,\sin(\theta) \,d\theta \frac{f_{l_p}(r_{Ap})\,g_{l_d}(r_d)}{r_d}u_d(r_{pn})
 \left[ Y^{l_p} (\theta) Y^{l_d} (\theta_d)\right]^l_0.
 \end{align}
\subsection{Neutron wavefunction}
The partial wave expansion of the Green's function (\ref{eq31}) for a given neutron energy  $\varepsilon$ can be written as
  \begin{equation}\label{eq65}
 G_l(r_{Bn},r_{Bn}')= \frac{f_l(k_n,r_{Bn<})g_l(k_n,r_{Bn>})}{k_n r_{Bn}r_{Bn}'},
 \end{equation}
 where $k_n=\sqrt{2 m_n \varepsilon}/\hbar$, and $f_l(k_n,r_{Bn})\, (g_l(k_n,r_{Bn}))$ is the regular (irregular) solution of the homogeneous equation
 \begin{align}\label{eq67}
 \left(-\frac{\hbar^2}{2m_n}\frac{\partial^2}{\partial r_{Bn}^2}+U_{Bn}(r_{Bn})+\frac{\hbar^2 l(l+1)}{2m_n r_{Bn}^2}-\varepsilon\right)\,\left\{f_l(k_n,r_{Bn}) ,g_l(k_n,r_{Bn})\right\}=0.
 \end{align}
The neutron wavefunction 
 \begin{equation}\label{eq91}
\psi_n(\mathbf{r}_{Bn};\mathbf{k}_p)= \sum_{l,m,l_p}\phi_{lml_p}(r_{Bn};\mathbf{k}_{p})Y^l_m(\theta_{Bn})Y^{l_p}_{-m} (\hat k_{p})/r_{Bn},
\end{equation}
can then be obtained with according to 
eq. (\ref{eq143}),    
\begin{align}\label{eq68}
\nonumber \phi_{lml_p}(r_{Bn},k_p)=&\int G_l(r_{Bn},r_{Bn}') \, \mathcal F_{lml_p}(r_{Bn}';k_p)\, r_{Bn}'^2dr_{Bn}'\\
\nonumber&=\frac{1}{k_n}\left(g_l(k_n,r_{Bn})\int_0^{r_{Bn}}f_l(k_n,r_{Bn}')\mathcal F_{lml_p}(r_{Bn}';k_p)\, r_{Bn}'dr_{Bn}'\right.\\
&\left.+f_l(k_n,r_{Bn})\int_{r_{Bn}}^{\infty}g_l(k_n,r_{Bn}') \mathcal F_{lml_p}(r_{Bn}';k_p)\, r_{Bn}'dr_{Bn}'\right).
\end{align}
 \begin{figure}
 \begin{center}
\includegraphics[width=8cm,angle=0]
{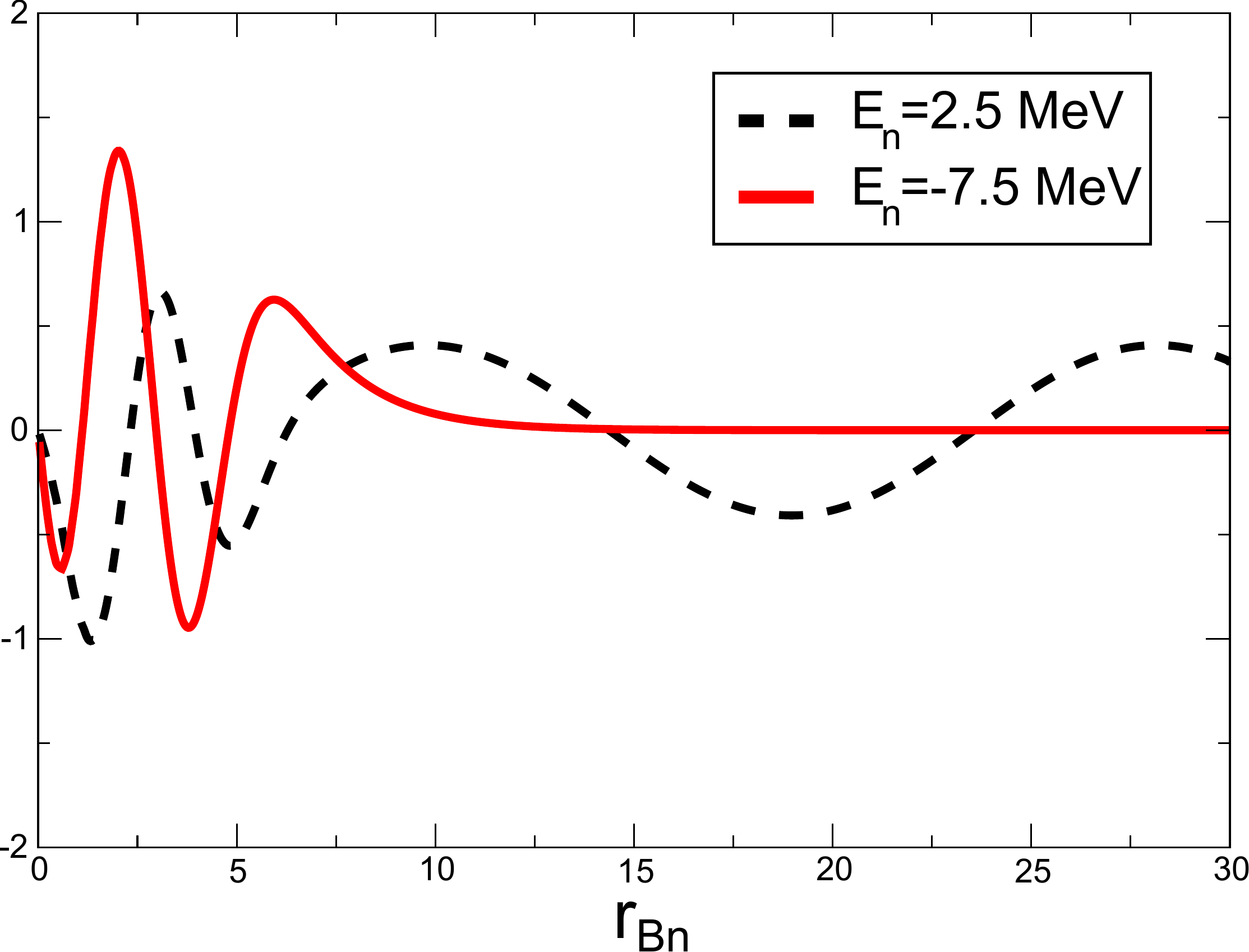}
\end{center}
\caption{Neutron partial wave coefficient   $\phi_{000}(r_{Bn})$ for $\varepsilon=2.5$ MeV (dashed black line)  and $\varepsilon=-7.5$ MeV  (red line).}\label{fig2}
\end{figure}
 \begin{figure}
 \begin{center}
\includegraphics[width=17cm]
{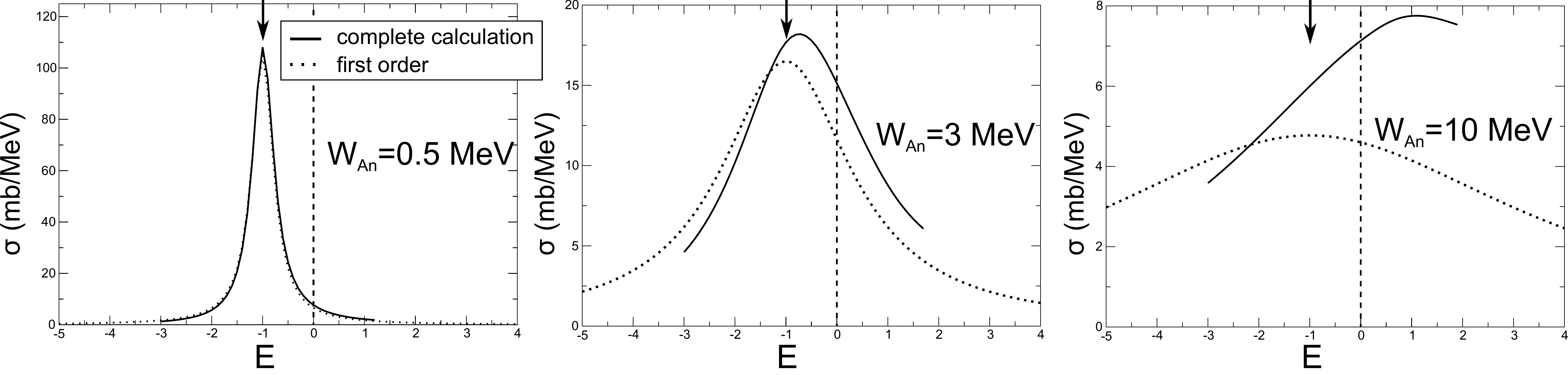}
 \end{center}
\caption{Non--elastic breakup cross section computed at neutron energies $E$ around a resonance $E_n=-1$ MeV. We   compare  the complete calculation (left side of eq. (\ref{eq124})) with the isolated--resonance, first--order approximation (right side of eq. (\ref{eq124})), for $W_{An}=0.5$ MeV, $W_{An}=3$ MeV and $W_{An}=10$ MeV. The  arrow indicates the value of the  eigenstate $E_n$ corresponding to the real part of the optical potential $U_{An}$, and the vertical dashed line is drawn at the neutron--emission threshold.}\label{fig3}
\end{figure}
\section{Results}\label{results}
It is important to note that the neutron wavefunctions (\ref{eq91}) are not eigenfunctions of a hermitian Hamiltonian, and can be associated with \textit{any} arbitrary energy $\varepsilon$, both positive and negative. In order to get the  physical wavefunctions, the corresponding boundary conditions have to be enforced by implementing them in the Green's function (\ref{eq65}). In order to do that,  we impose $\lim_{r_{An}\rightarrow 0}f_l(k_n,r_{An})=0$ for the regular solution. At large distances the boundary condition of course depends on whether the energy $\varepsilon$ is positive or negative. For scattering neutron states (positive $\varepsilon$),
 \begin{align}\label{eq70p}
\lim_{r_{An}\rightarrow \infty}g_l(k_n,r_{An})\rightarrow e^{i(k_nr_{An}-\frac{l\pi}{2})}, 
 \end{align}
 while for final neutron bound states (negative $\varepsilon$),
  \begin{align}\label{eq70n}
 \lim_{r_{An}\rightarrow \infty}g_l(k_n,r_{An})\rightarrow e^{-(\kappa_nr_{An})}, 
  \end{align}
with  $\kappa_n=\sqrt{-2m_n\varepsilon}/\hbar$. This last, somewhat less standard, condition can be implemented by integrating inwards numerically a function with the boundary  condition 
  \begin{align}\label{eq92}
g_l(k_n,R_\infty-h)=\frac{g_l(k_n,R_\infty)}{(1-\kappa_nh)},
  \end{align}
where $R_\infty$ is a large value of the radius and $h$ is the numerical integration step, chosen such that $\kappa_nh\ll1$.
We can thus use eq. (\ref{eq68}) to obtain the neutron wavefunction for arbitrary positive and negative energies. As an example, we show in Fig. \ref{fig2} the wavefunction $\phi_{000}$ for $\varepsilon=2.5$ MeV and $\varepsilon=-7.5$ MeV. The neutron wavefunction (\ref{eq91}) and the non--orthogonality term (\ref{eq71}) can then be used in (\ref{eq58}) to obtain the non--elastic breakup cross section. If the final neutron energy is negative, the capture of the neutron in a region in which the imaginary part $W_{An}$ of the optical is small is related to the direct transfer to a sharp bound state. Actually, it can be shown (see \cite{Potel:15b})  that, in first order of $\langle W_{An} \rangle\equiv\langle \phi_n| W_{An}|\phi_n \rangle$,  there is a simple relationship between the cross section for the capture of a neutron in a bound state of finite width and the cross section for the direct transfer to the corresponding zero--width bound state. Assuming  that the 
one--neutron transfer DWBA amplitude 
\begin{align}\label{eq150}
T^{\rm (1NT)}_n=\int \phi^*_n\left(\,\chi_f^{(-)}\big |V_{prior}\big |\phi_d\,\chi_i\right\rangle d\mathbf r'_{An}
\end{align}
to the single--particle state $\phi_n$ of the target--neutron residual nucleus is  constant in an energy range of the order of $\Gamma_n=2\langle  W_{An} \rangle$, we have
 \begin{align}\label{eq124}
 \left.\frac{d^2\sigma}{d\Omega_pdE_p}(E,\Omega)\right]^{NEB}\approx\frac{1}{2\pi}\frac{\Gamma_n}{\left(E_n-E\right)^2+\Gamma_n^2/4}\frac{d\sigma_n}{d\Omega}(\Omega),
 \end{align}
where $\frac{d\sigma_n}{d\Omega}$ is the direct transfer differential cross section to the $n$th eigenstate of the real potential. For the above approximation to be valid, $\langle  W_{An} \rangle$ needs to be small, and the distance $\Delta E$ between the resonance $E_n$ and te closest one has to be big enough ($\Delta E\gg \Gamma_n$). In particular, the latter condition is hardly verified if the resonance is too close (i.e., within a distance of the order or smaller than $\Gamma_n$) to the continuum (neutron emission threshold). In Fig. \ref{fig3} we compare the first order approximation with the exact calculation for energies close to a resonance 1 MeV away from the neutron emission threshold, for three different values of $W_{An}$.
\section{Conclusions}
We have presented  a formalism for inclusive deuteron--induced reactions, in which the final neutron--target system is left in an arbitrary state characterized by its energy, angular momentum and parity. The general derivation of the expression of the non--elastic breakup eq. (\ref{eq58}) is given elsewhere (\cite{Potel:15b}), as well as the comparison with experimental results. In this paper we have focused our attention in the final neutron wavefunction, presenting an explicit expression that can be computed numerically (see eqs. (\ref{eq91}) and (\ref{eq68})). If the target--neutron interaction is modeled with an optical  potential $U_{An}$ with a non--zero imaginary part $W_{An}$, the negative--energy part of the neutron spectrum is no longer discrete. Instead of being composed by sharp single--particle states, the neutron can have any continuous value of the energy above the Fermi energy. This continuous spectrum exhibit a resonant behavior around a discrete set of energies, with widths naturally related with the value of $W_{An}$. We show that, in the limit in which $W_{An}$ is small compared to the real part of $U_{An}$ and to the distance between resonances, the set of discrete resonances can be related to the discrete single--particle spectrum, i.e., the set of eigenvalues corresponding to the real part of $U_{An}$. Moreover, the energy--integrated cross section around a resonance gives the direct DWBA transfer cross section to that particular state. 
\begin{center}
\textbf{ACKNOWLEDGEMENT}
\end{center}
This work was supported by the National Science
Foundation under Grant No. PHY-1403906,   and the Department
of Energy, Office of Science, Office of Nuclear Physics under award No. DE-FG52-08NA28552,
and by Lawrence Livermore National Laboratory under Contract DE-AC52-07NA27344.

 \bibliographystyle{ieeetr}

\begin{thebibliography}{10}

\bibitem{Kerman:79}
A.~Kerman and K.~McVoy, ``Fluctuations in two-step reactions through
  doorways,'' {\em Annals of Physics}, vol.~122, no.~1, p.~197, 1979.

\bibitem{Udagawa:80}
T.~Udagawa and T.~Tamura, ``Breakup-fusion description of massive transfer
  reactions with emission of fast light particles,'' {\em Phys. Rev. Lett.},
  vol.~45, p.~1311, 1980.

\bibitem{Austern:81}
N.~Austern and C.~M. Vincent, ``Inclusive breakup reactions,'' {\em Phys. Rev.
  C}, vol.~23, p.~1847, 1981.

\bibitem{Li:84}
X.~H. Li, T.~Udagawa, and T.~Tamura, ``Assessment of approximations made in
  breakup-fusion descriptions,'' {\em Phys. Rev. C}, vol.~30, p.~1895, 1984.

\bibitem{Udagawa:86}
T.~Udagawa and T.~Tamura, ``Formulation of elastic and inelastic breakup-fusion
  reactions,'' {\em Phys. Rev. C}, vol.~33, p.~494, 1986.

\bibitem{Ichimura:85}
M.~Ichimura, N.~Austern, and C.~M. Vincent, ``Equivalence of post and prior sum
  rules for inclusive breakup reactions,'' {\em Phys. Rev. C}, vol.~32, p.~431,
  1985.

\bibitem{Ichimura:90}
M.~Ichimura, ``Relation among theories of inclusive breakup reactions,'' {\em
  Phys. Rev. C}, vol.~41, p.~834, 1990.

\bibitem{Hussein:85}
M.~Hussein and K.~McVoy, ``{Inclusive projectile fragmentation in the spectator
  model},'' {\em Nuclear Physics A}, vol.~445, p.~124, 1985.

\bibitem{Hussein:89}
M.~S. Hussein and R.~Mastroleo, ``{Glauber calculation of heavy--ion and
  light-ion inclusive break-up cross sections},'' {\em Nuclear physics A},
  vol.~491, p.~468, 1989.

\bibitem{Potel:15b}
G.~Potel, F.~M. Nunes, and I.~J. Thompson, ``{Establishing a theory for
  neuteron--induced surrogate reactions},'' {\em {Phys. Rev. C}}, vol.~92,
  p.~034611, 2015.

\bibitem{Lei:15}
J.~Lei and A.~Moro, ``Revisiting closed-form formulae for inclusive breakup,''
  {\em Physical Review C}, 2015.
\newblock To be published.

\bibitem{Carlson:15}
B.~V. Carlson, R.~Capote, and M.~Sin, ``{Elastic and inelastic breakup of
  deuterons with energy below 100 MeV},'' {\em arXiv:1508.01466 [nucl-th]},
  2015.

\bibitem{Escher:10}
J.~Escher and F.~Dietrich, ``{Cross sections for neutron capture from surrogate
  measurements: An examination of Weisskopf--Ewing and ratio approximations},''
  {\em Phys. Rev. C}, vol.~81, p.~024612, 2010.

\bibitem{Scielzo:10}
N.~D. Scielzo, J.~E. Escher, J.~M. Allmond, M.~S. Basunia, C.~W. Beausang,
  L.~A. Bernstein, D.~L. Bleuel, J.~T. Burke, R.~M. Clark, F.~S. Dietrich,
  P.~Fallon, J.~Gibelin, B.~L. Goldblum, S.~R. Lesher, M.~A. McMahan, E.~B.
  Norman, L.~Phair, E.~Rodriquez-Vieitez, S.~A. Sheets, I.~J. Thompson, and
  M.~Wiedeking, ``{Measurement of $\gamma$-emission branching ratios for
  $^{154,156,158}\mathrm{Gd}$ compound nuclei: Tests of surrogate nuclear
  reaction approximations for $(n,\gamma$) cross sections},'' {\em Phys. Rev.
  C}, vol.~81, p.~034608, 2010.

\bibitem{Escher:12}
J.~Escher, J.~Burke, F.~Dietrich, N.~Scielzo, I.~Thompson, and W.~Younes,
  ``{Compound--nuclear reaction cross sections from surrogate measurements},''
  {\em Rev. Mod. Phys.}, vol.~84, p.~353, 2012.

\bibitem{Hughes:12}
R.~O. Hughes, C.~W. Beausang, T.~J. Ross, J.~T. Burke, N.~D. Scielzo, M.~S.
  Basunia, C.~M. Campbell, R.~J. Casperson, H.~L. Crawford, J.~E. Escher,
  J.~Munson, L.~W. Phair, and J.~J. Ressler, ``{Utilizing ($p$,$d$) and
  ($p$,$t$) reactions to obtain ($n$,$f$) cross sections in uranium nuclei via
  the surrogate-ratio method},'' {\em Phys. Rev. C}, vol.~85, p.~024613, 2012.

\bibitem{Ross:12}
T.~J. Ross, C.~W. Beausang, R.~O. Hughes, J.~M. Allmond, C.~T. Angell, M.~S.
  Basunia, D.~L. Bleuel, J.~T. Burke, R.~J. Casperson, J.~E. Escher, P.~Fallon,
  R.~Hatarik, J.~Munson, S.~Paschalis, M.~Petri, L.~Phair, J.~J. Ressler, N.~D.
  Scielzo, and I.~J. Thompson, ``Measurement of the entry-spin distribution
  imparted to the high excitation continuum region of gadolinium nuclei via
  ($p$,$d$) and ($p$,$t$) reactions,'' {\em Phys. Rev. C}, vol.~85, p.~051304,
  2012.

\bibitem{Carlson:14}
B.~Carlson, J.~Escher, and M.~Hussein, ``{Theoretical descriptions of
  compound--nuclear reactions: open problems and challenges},'' {\em J. Phys.
  G}, vol.~41, p.~094003, 2014.

\end{thebibliography}
\end{document}